# Scale-free network of earthquakes

Sumiyoshi Abe[1] and Norikazu Suzuki[2]

[1]*Institute of Physics, University of Tsukuba, Ibaraki 305-8571, Japan*
[2]*College of Science and Technology, Nihon University, Chiba 274-8501, Japan*

**Abstract.** The district of southern California and Japan are divided into small cubic cells, each of which is regarded as a vertex of a graph if earthquakes occur therein. Two successive earthquakes define an edge and a loop, which replace the complex fault-fault interaction. In this way, the seismic data are mapped to a random graph. It is discovered that an evolving random graph associated with earthquakes behaves as a scale-free network of the Barabási-Albert type. The distributions of connectivities in the graphs thus constructed are found to decay as a power law, showing a novel feature of earthquake as a complex critical phenomenon. This result can be interpreted in view of the facts that frequency of earthquakes with large values of moment also decays as a power law (the Gutenberg-Richter law) and aftershocks associated with a mainshock tend to return to the locus of the mainshock, contributing to the large degree of connectivity of the vertex of the mainshock. It is also found that the exponent of the distribution of connectivities is characteristic for a plate under investigation.

There is growing interest in the earthquake phenomenon from the viewpoint of science of complex systems [1-7]. Though seismicity is characterized by remarkably rich phenomenology, some of the known empirical laws are rather simple. The Omori law [8] for temporal distribution of aftershocks and the Gutenberg-Richter law [9] for the relationship between frequency and magnitude are regarded as classical examples.

Recently, we have studied the earthquake phenomenon from the perspective of nonextensive statistical mechanics [10-13]. Nonextensive statistical mechanics is a new theory, which is based on the Tsallis nonadditive entropy with the entropic index $q\,(>0)$ [14] and is currently under vital investigation with anticipation as a unified framework for understanding the statistical properties of complex systems in their quasi-equilibrium states. The distribution, which optimizes the Tsallis entropy under appropriate constraints, is termed the "$q$-exponential distribution" [15]. [The associated "$q$-exponential function" of $x$ is defined by $e_q(x) = (1+(1-q)x)_+^{1/(1-q)}$ with the notation $(a)_+ \equiv \max\{0, a\}$, and converges to the ordinary exponential function in the limit $q \to 1$. See also the discussion after eq. (2) below.] We have discovered the remarkable relevance of nonextensive statistical mechanics to spatio-temporal complexity of the earthquake phenomenon. We have found by analyzing the seismic data the following results. For the spatial structure, the distance between successive earthquakes obeys the $q$-exponential distribution with $q$ less than unity [16]. On the other hand, for the temporal structure, the waiting time distribution associated with successive earthquakes undergoes over a series of transitions between stationary states, each of which again follows the $q$-exponential distribution with $q$ larger than unity [17], which is equivalent to the Zipf-Mandelbrot distribution [18].

Now, in modern statistical mechanics of complex systems, the idea of evolving random graphs is receiving great attention. The research in this direction has been initiated by the works of Watts and Strogatz [19] on small worlds and Barabási and Albert [20] on scale-free evolving random networks. One of its main subject is to understand topology of the complex network structure. Examples include collaboration of actors, pattern of citation of scientific papers, metabolic networks, social networks, World Wide Web, electrical power grids etc. [20-23]. Each of these systems is mapped to a random graph, in which a member (e.g., HTML in the case of World Wide Web) is identified with a vertex and the complex interaction between the members (e.g., pointing from one HTML to another) is replaced by an edge. An interesting observation common in these apparently different systems is that the evolving networks can self-organize into the states described by the connectivity distributions that decay as a power law for large values of connectivity. The number of vertices and the pattern of edges change in time in a probabilistic way. This is a novel approach to complex systems with the lack of detailed knowledge about microscopic interactions between members.

The crucial assumption in the Barabási-Albert model is that a newly created vertex is connected to the vertex $i$ with connectivity $k_i$ with probability

$$\Pi(k_i) = \frac{k_i + 1}{\sum_j (k_j + 1)}. \tag{1}$$

That is, the new vertex tends to be connected to a vertex with larger connectivity. In [24], Albert and Barabási have discussed an solvable model of an evolving random graph and have presented the analytic expression for the distribution of connectivities, $P(k)$, corresponding to the number of vertices with $k$ edges, in the continuum limit. This solution is actually the Zipf-Mandelbrot-type distribution

$$P(k) \sim (k + k_0)^{-\gamma} \tag{2}$$

with $\gamma > 1$, which clearly decays as a power law. In some interesting examples such as actor collaboration, World Wide Web, and electrical power grids, the value of $\gamma$ ranges between 2 and 3 [22]. We wish to point out that, using the $q$-exponential function, $P(k)$ in eq. (2) is rewritten as $P(k) \sim e_q(-k/\kappa)$, where $q = 1 + 1/\gamma$ and $\kappa = (q-1)k_0$. This implies that the Albert-Barabási solution optimizes the Tsallis entropy [14], $S_q[P] = (1-q)^{-1}\left(\int dk\, P^q(k) - 1\right)$, under the appropriate constraints on the normalization of $P(k)$ and the average number of edges [10-13,15] in the continuum limit.

Now, as mentioned earlier, we have recently discovered [16,17] that the $q$-

exponential distributions in nonextensive statistical mechanics are profoundly relevant to the complex spatio-temporal properties of the earthquake phenomenon. To understand the physics underlying such relevance deeper, it is important to examine if there also exists the structure of scale-free networks in seismicity. In this Letter, we present an evidence supporting this viewpoint. We define the evolving network associated with earthquakes, referred to here as the earthquake network, and apply this concept to the seismic data taken in southern California and Japan. Remarkably, we shall see that the evolving networks in these two areas are in fact scale-free. *Analysis of the seismic data shows that aftershocks associated with a mainshock tend to return to the locus of the mainshock geographically* and a stronger mainshock tends to have the larger degree of connectivity contributed by more aftershocks. Thus, the present result can naturally be interpreted if combined with the Gutenberg-Richter law, which states that frequency of earthquakes with large values of moment also decays as a power law.

Our proposal for constructing the earthquake network is as follows. The geographical region under consideration is divided into a lot of small cubic cells. Each cell, in which events with any magnitudes occurred, is identified with a vertex. Two successive events define an edge between two vertices. In other words, the complex fault-fault interaction is replaced by this edge. Two vertices may sometimes coincide with each other (i.e., successive events occurring in the same cell), forming a loop.

We have constructed the earthquake networks in the district of southern California and Japan by introducing two cell sizes, 10km×10km×10km and 5km×5km×5km. (There are no *a priori* operational rules to fix the cell size, and this is the reason why we have examined these two cases.) In these constructions, we have analyzed the earthquake catalogs: one is made available by the Southern California Earthquake Data Center (http://www.scecdc.scec.org/ catalogs.html) covering the region 29°15.25'N–38°49.02'N latitude and 113°09.00'W–122°23.55'W longitude with the maximal depth (of the foci of the observed earthquakes) 57.88km in the period between 00:25:8.58 on January 1, 1984 and 15:23:54.73 on December 31, 2002, and the other is by the Japan University Network Earthquake Catalog (http://kea.eri.u-tokyo.ac.jp/CATALOG/junec/monthly. html) covering the region 25.730°N–47.831°N latitude and 126.433°E–148.000°E longitude with the maximal depth 599.9km in the period between 01:14:57.63 on January 1, 1993 and 20:54:38.95 on December 31, 1998. The total numbers of the events are 375887 and 123390, respectively. (We have limited ourselves to the above period in Japan since before 1993 the number of the observed data per year turned out to be about half of the later period. We also note that the data in southern California contain the events with arbitrary magnitude, whereas the Japanese data does not contain the events smaller than magnitude 2, unfortunately.)

In Figs. 1 and 2, we present the plots of the distributions of connectivities in southern California and Japan, respectively. The values of the exponent $\gamma$ in Japan are large compared to those in southern California. Small earthquakes (which are not contained in the Japanese data) tend to contribute to large values of connectivities, making the exponent smaller. Since the number of vertices with large degrees of connectivity decreases if the cell size becomes smaller, the value of $\gamma$ is larger for smaller cells, in general. Though we do not explicitly present here, actually we have also performed analysis of the data in southern California larger than magnitude 2 (as in the Japanese data) in the case of the size of the cell $5\,\text{km} \times 5\,\text{km} \times 5\,\text{km}$. The result found is $\gamma = 1.75$, which is still much smaller than the corresponding Japanese value, $\gamma = 2.50$.

Finally, we would like to report the following further two discoveries on time evolution of the earthquake network. (i) The factor, $k_0$, appearing in eq. (2) is found to change in time, in contrast to the Albert-Barabási solution given in [24]. This is shown in Fig. 3, where the data taken in southern California are employed. Clearly, monotonic increase of the value of $k_0$ is observed there. (ii) *The value of the exponent, $\gamma$, is ascertained to remain constant in time according to evolution of the earthquake network.* The fact that $\gamma$ is constant in time indicates that this exponent may be characteristic for a plate under investigation. The significant difference between the values of $\gamma$ in southern California and Japan mentioned above also supports this viewpoint.

In conclusion, we have constructed the earthquake networks in the district of southern California and Japan and have analyzed the seismic data from the novel viewpoint of growing random graphs. We have found that the earthquake networks possess the scale-free nature of the Barabási-Albert type in their distributions of connectivities and, thus, have presented a novel feature of earthquake as a complex critical phenomenon. We have also observed that the exponent, $\gamma$, of the distribution may characterize the physical property of a plate under investigation. The scale-free nature of the earthquake network may accept the following natural interpretation. Frequency of earthquakes with large values of moment decays as a power law due to the Gutenberg-Richter law, on the one hand, and analysis of the seismic data, on the other hand, shows that aftershocks associated with a mainshock tend to return to the locus of the mainshock geographically and therefore contribute to the large degree of connectivity of the vertex of the mainshock. This may be the origin of the scale-free nature of the earthquake network. As a future problem, it is of theoretical interest to explore possibility of generalizing the Albert-Barabási discussion [24] to incorporate the time-dependent $k_0$ factor.

**References**

[1] P. Bak and C. Tang, J. Geophys. Res. **94**, 15635 (1989).

[2] Z. Olami , H. J. S. Feder, and K. Christensen, Phys. Rev. Lett. **68**, 1244 (1992).

[3] Y. Huang, H. Saleur, C. Sammis, and D. Sornette, Europhys. Lett **41**, 43 (1998).

[4] A. Sornette and D. Sornette, Geophys. Res. Lett. **26**, 1981 (1999).

[5] Y. Huang, A. Johansen, M. W. Lee, H. Saleur, and D. Sornette,

J. Geophys. Res. **105** (B11), 25451 (2000).

[6] P. Bak, K. Christensen, L. Danon, and T. Scanlon,

Phys. Rev. Lett. **88**, 178501 (2002).

[7] S. Lise and M. Paczuski, Phys. Rev. Lett. **88**, 228301 (2002).

[8] F. Omori, J. Coll. Sci. Imp. Univ. Tokyo **7**, 111 (1894).

[9] B. Gutenberg and C. F. Richter, Bull. Seism. Soc. Am. **34**, 185 (1944).

[10] C. Tsallis, R. S. Mendes, and A. R. Plastino, Physics A **261**, 534 (1998).

[11] *Nonextensive Statistical Mechanics and Its Applications*, edited by

S. Abe and Y. Okamoto (Springer-Verlag, Heidelberg, 2001).

[12] Special issue of Physica A **305** (2002), edited by G. Kaniadakis, M. Lissia,

and A. Rapisarda.

[13] *Nonextensive Entropy—Interdisciplinary Applications*, edited by

M. Gell-Mann and C. Tsallis (Oxford university Press, Oxford), in press.

[14] C. Tsallis, J. Stat. Phys. **52**, 479 (1988).

[15] S. Abe, Phys. Rev. E **66**, 046134 (2002).

[16] S. Abe and N. Suzuki, J. Geophys. Res. **108** (B2), 2113 (2003).

[17] S. Abe and N. Suzuki, Zipf-Mandelbrot law for time intervals of earthquakes,

e-print cond-mat/0208344.

[18] B. B. Mandelbrot, *The Fractal Geometry of Nature* (Freeman, San Francisco, 1983).

[19] D. J. Watts and S. H. Strogatz, Nature (London) **393**, 440 (1998).

[20] A.-L. Barabási and R. Albert, Science **286**, 509 (1999).

[21] A.-L. Barabási, Physics World **14**, 33 (July, 2001).

[22] R. Albert and A.-L. Barabási, Rev. Mod. Phys. **74**, 47 (2002).

[23] S. N. Dorogovtsev and J. F. F. Mendes, *Evolution of Networks* (Oxford University Press, Oxford, 2003).

[24] R. Albert and A.-L. Barabási, Phys. Rev. Lett. **85**, 5234 (2000).

# Figure Captions

Fig. 1. The log-log plots of the distributions of connectivities in southern California. All quantities are dimensionless. The dots represent the observed data and the solid lines are drawn by using eq. (2). (a) The size of the cell $10\text{km} \times 10\text{km} \times 10\text{km}$, $\gamma = 1.36$, and $k_0 = 1.65$. (b) The size of the cell $5\text{km} \times 5\text{km} \times 5\text{km}$, $\gamma = 1.61$, and $k_0 = 2.04$.

Fig. 2. The log-log plots of the distributions of connectivities in Japan. All quantities are dimensionless. The dots represent the observed data and the solid lines are drawn by using eq. (2). (a) The size of the cell $10\text{km} \times 10\text{km} \times 10\text{km}$, $\gamma = 2.22$, and $k_0 = 1.71$. (b) The size of the cell $5\text{km} \times 5\text{km} \times 5\text{km}$, $\gamma = 2.50$, and $k_0 = 0.79$.

Fig. 3. Change of the value of $k_0$ in southern California according to evolution of the earthquake network. Here, the size of the cell employed is $5\text{km} \times 5\text{km} \times 5\text{km}$. The value of the exponent remains constant: $\gamma = 1.61$.

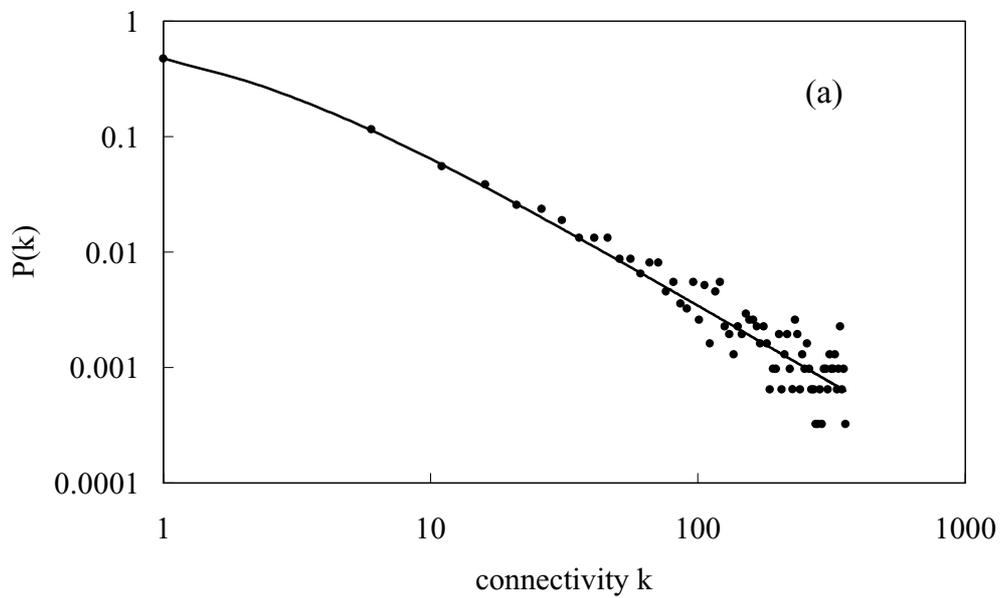

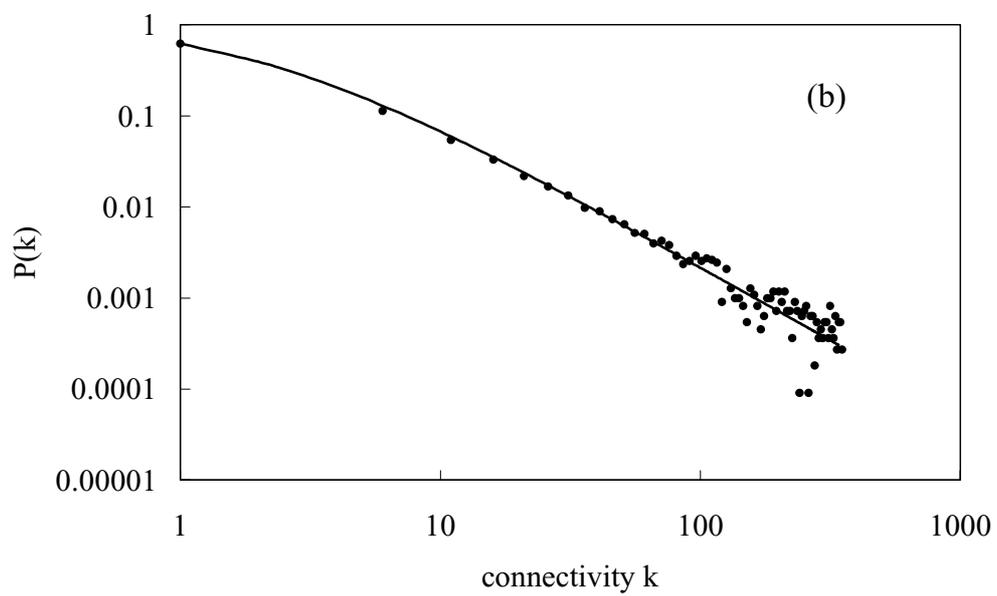

Fig. 1

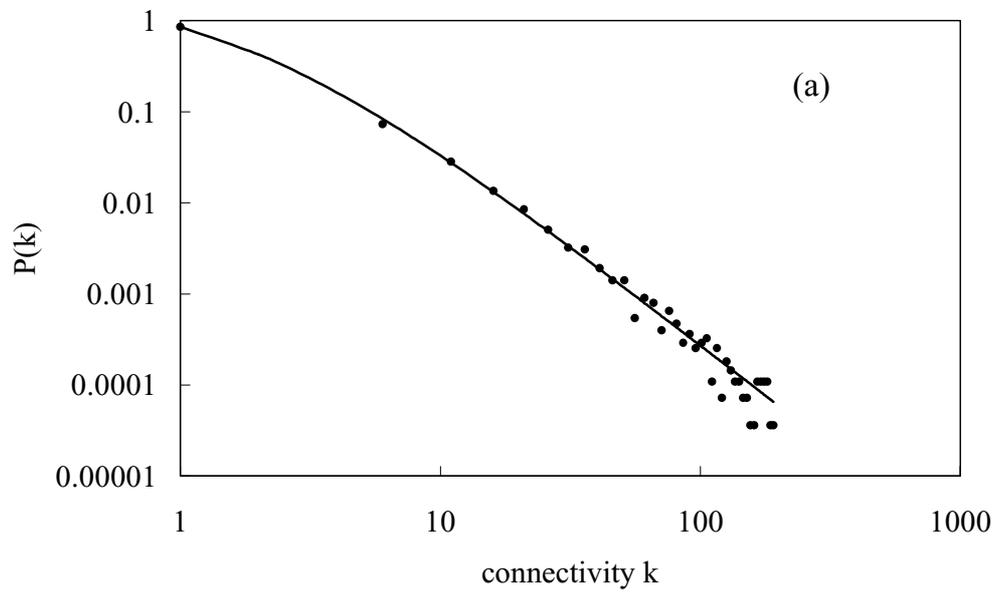

(a)

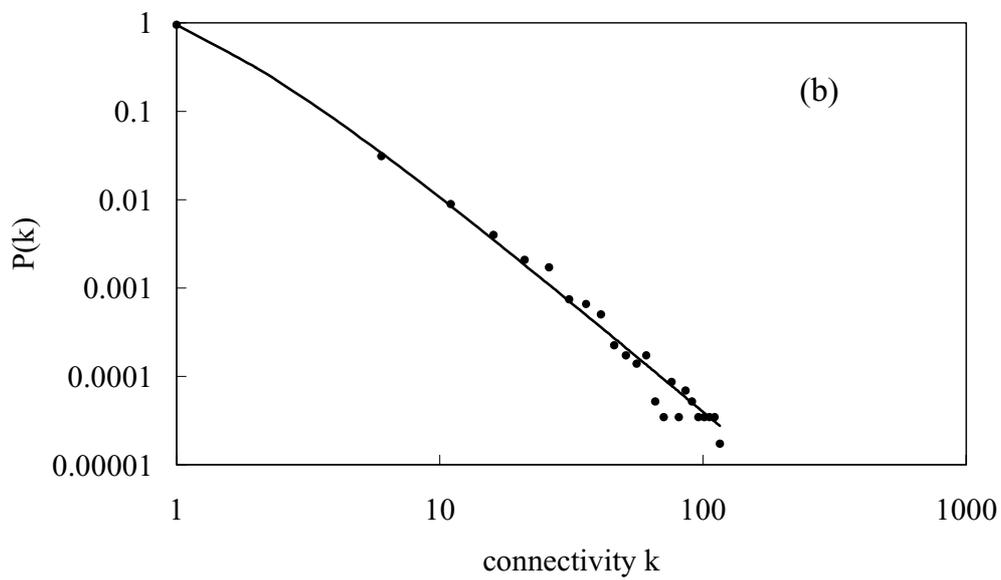

(b)

Fig. 2

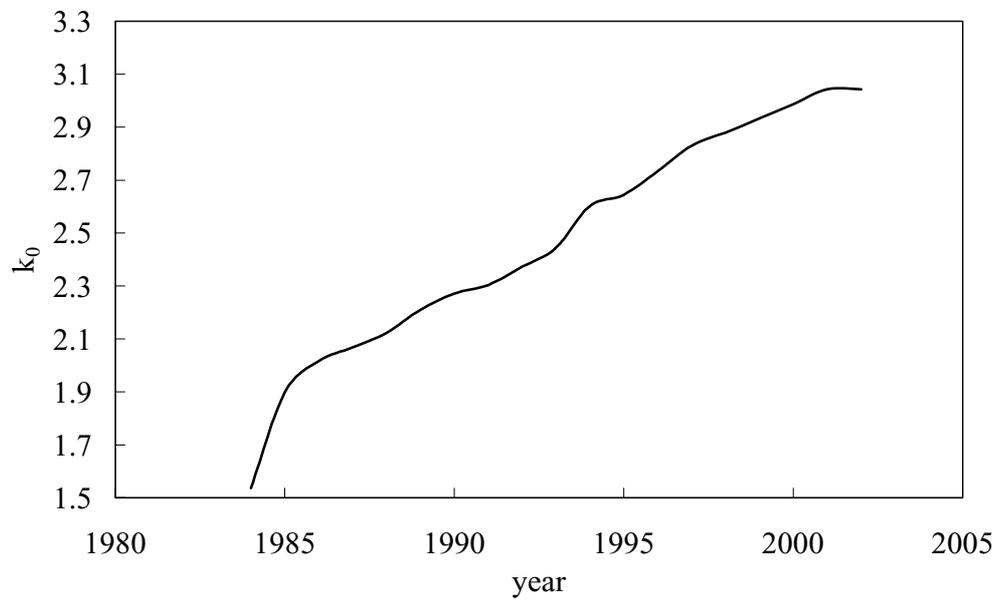

Fig. 3